\documentclass[%
preprint,
superscriptaddress,
showpacs,preprintnumbers,
 amsmath,amssymb,
 aps,
 prl,
]{revtex4-1}

\usepackage{graphicx}
\usepackage{dcolumn}
\usepackage{bm}
\usepackage{color}
\usepackage{subfigure}

\begin{document}

\title{Turbulence of Dilute Polymer Solution}

\author{Heng-Dong Xi}
\affiliation{Max-Planck Institute for Dynamics and Self-Organization (MPIDS), 37077 G\"{o}ttingen, Germany }
\author{Eberhard Bodenschatz}
\affiliation{Max-Planck Institute for Dynamics and Self-Organization (MPIDS), 37077 G\"{o}ttingen, Germany }
\affiliation{Institute for Nonlinear Dynamics, University of G\"{o}ttingen, 37077 G\"{o}ttingen, Germany}
\affiliation{Laboratory of Atomic and Solid State Physics and Sibley School of Mechanical and Aerospace Engineering, Cornell University, Ithaca, NY 14853, USA}

\author{Haitao Xu}
\email{Email for correspondance: haitao.xu@ds.mpg.de}
\affiliation{Max-Planck Institute for Dynamics and Self-Organization (MPIDS), 37077 G\"{o}ttingen, Germany }
\date{\today}

%


\maketitle

\textbf{In fully developed three dimensional fluid turbulence the fluctuating energy is supplied at large scales, cascades through intermediate scales, and dissipates at small scales.  It is the hallmark of turbulence that for intermediate scales, in the so called inertial range,  the average energy flux is constant and independent of viscosity \cite{richardson:1922,K41,frisch:1995}.  One very important question is how this range is altered, when an additional agent that can also transport energy is added to the fluid.  Long-chain polymers dissolved at very small concentrations in the fluid are such an agent \cite{groisman:2000, procaccia:2008}. Based on prior work by de Gennes and Tabor \cite{tabor:1986,degennes:1986} we introduce a theory that balances the energy flux  through the turbulent cascade with that of the energy flux into the elastic degrees of freedom of the dilute long-chain polymer solution. We propose a refined elastic length scale, $r_\varepsilon$, which describes the effect of polymer elasticity on the turbulence energy cascade.  Our experimental results agree excellently with this new energy flux balance theory.  }

To date, the only relevant theory on the interaction between polymers and turbulence cascade is the ``energy balance theory"  by de Gennes and Tabor~\cite{tabor:1986,degennes:1986} as summarized recently by Sreenivasan and White  \cite{sreenivasan:2000}. It states that the turbulence energy cascade is essentially unaltered down to a scale at which the energy stored in  the elastic degrees of freedom of the polymer is equal to the kinetic energy of the flow. So far experiments on fully developed three dimensional turbulence  do not convincingly support this theory ~\cite{ouellette:2009}. Here we argue that this may not be surprising as it is the turbulent flux of energy and not the energy itself that determines the inertial range properties of turbulence.  

In this paper, we provide what may be called an ``energy flux balance theory''. In our theory, the turbulent energy flux through the cascade (or the turbulent energy transfer rate from scale to scale) is gradually reduced by the energy transfer through stretching and recoiling of the polymer chains, with the elastic energy flux becoming dominant at small scales. For the sake of making progress, we assume that the balance of elastic and turbulent energy occurs in average at one length scale. This of course is a crude assumption that does not hold in detail as the turbulent energy flux is known be intermittent  \cite{frisch:1995}. Nevertheless, we may expect that it captures the main features, just as Kolomogorov's  1941 (K41) theory \cite{K41} does for pure fluids.  Here we show, that the energy flux balance theory is supported quantitatively by our experimental data measured over a wide range of parameters in fully developed turbulence. Our approach may also be applied to and provide new perspectives into turbulence in other complex flows involving two or more nonlinear mechanisms, like for example,  turbulence in conducting fluids, plasmas or quantum fluids.

For stationary turbulence in Newtonian fluids, the kinetic energy of the turbulent motion per unit mass is supplied at a rate of $\varepsilon_I$ at the forcing scales: length scale $L$ and time scale $T_L$, which are related by $\varepsilon_I \approx u^2/T_L \approx u^3/L$, where $u$ is the root mean square of the fluctuating velocity. As the scale decreases, the viscous effects become more and more important.  It is well known, that  at the Kolmogorov scale the viscous forces dominate and dissipate kinetic energy to heat at a rate of $\varepsilon_D$, which gives naturally the length scale $\eta \equiv (\nu^3/\varepsilon_D)^{1/4}$ and the time scale $\tau_\eta \equiv (\nu / \varepsilon_D)^{1/2}$, where $\nu$ is the kinematic viscosity of the fluid. The Reynolds number $Re \equiv uL/\nu$, a parameter to characterize the intensity of the turbulence, measures the separation of scales: $L/\eta \sim Re^{3/4}$ and $T_L/\tau_\eta \sim Re^{1/2}$. For $Re \gg 1$, the forcing scales ($L$ and $T_L$) are widely separated from the dissipative scales ($\eta$ and $\tau_\eta$). The K41 theory~\cite{K41} then states that for intermediate scales $\eta \ll r \ll L$ or $\tau_\eta \ll \tau \ll T_L$, the energy is transferred down to smaller scales without loss. This immediately leads to the conclusion that $\varepsilon_I = \varepsilon_T(r) = \varepsilon_D = \varepsilon$ is a constant. Here we used the notation $\varepsilon_T(r)$ to emphasize that the energy transfer rate depends on scale $r$. This local energy transfer rate can be estimated as $\varepsilon_T(r) = {u_r}^2 / \tau_r$, where $u_r$ and $\tau_r$ are the characteristic velocity and time at scale $r$ ($u_r$ is often related to the velocity differences at scale $r$: $u_r \sim | u(x+r) - u(x)|$). The results from K41 theory is $u_r \sim (\varepsilon r)^{1/3}$ and $\tau_r \sim (r^2/\varepsilon)^{1/3}$, which is consistent with $\varepsilon_T(r) = \varepsilon$ for $\eta < r < L$.

This elegant picture of the energy cascade is changed when flexible long-chain polymers are dissolved in the fluid. For simplicity, one may regard  a single polymer chain as an entropic spring that is constantly stretching and coiling back in the flow~\cite{smith:1999,tabor:1986,degennes:1986} . The turbulence fluctuations at different scales contribute unequally to the stretching of the polymer chain. In particular, Lumley concluded that only those fluctuations with time scale $\tau_r \lesssim \tau_p$ can stretch the polymer chain ~\cite{lumley:1973}, where $\tau_p$ is the entropic viscous relaxation time of the polymer chain. This ``time criterion'', which is essentially the same as requiring that the scale-dependent Weissenberg number $Wi_r = \tau_p / \tau_r \gtrsim 1$, thus defines the Lumley scale $r^* \equiv (\varepsilon {\tau_p}^3)^{1/2}$ (see e.g. ~\cite{procaccia:2008}). The physical meaning of $r^*$ is that below this scale the local fluid deformation would be strong enough to stretch polymers.  

\begin{figure} [htb]
\includegraphics[width=1\columnwidth]{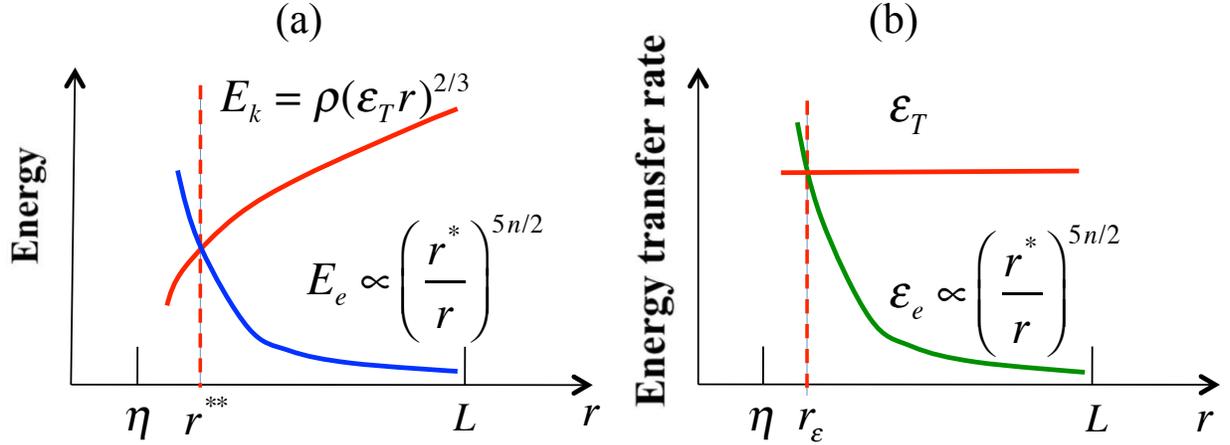}
\caption{\label{fig:balances}  (a) A sketch showing the balance between turbulent fluctuation energy and the polymer elastic energy, which gives the scale $r^{**}$~\cite{tabor:1986,degennes:1986}; (b) Similar sketch for the balance between the turbulent energy flux and the elastic energy flux, which determines the scale $r_{\varepsilon}$. }
\end{figure}

Please note, that the Lumley scale  can only  tell us on whether a single polymer chain can be stretched by turbulence or not and thus cannot address how polymers dissolved at a certain concentration will affect the flow. This was addressed by the ``energy balance theory'' proposed by Tabor \& de Gennes. By analogy to polymers in a linearly stretching field (see Methods), Tabor \& de Gennes~\cite{tabor:1986} suggested that the polymer elastic energy per volume is $E_e(r) \sim c_pkT(r^*/r)^{5n/2}$ for $r \ll r^*$, where $c_p$ is the number of polymer chains per unit volume, $k$ is the Boltzmann constant, $T$ is the temperature of the fluid, and $n$ is an unknown exponent that is related to the average stretching dimensions of the local flow field. De Gennes~\cite{degennes:1986} further argued that the turbulent energy cascade will be truncated below a scale $r^{**}$ at which the polymer elastic energy balances the kinetic energy of the turbulent fluctuations (see also Figure~\ref{fig:balances}(a)):
\begin{equation} \label{eqn:energy_balance}
\rho u_{r^{**}}^2 = \rho (\varepsilon_T r^{**})^{2/3} = c_pkT(\frac{r^*}{r^{**}})^{5n/2}.
\end{equation}
This gives 
\begin{equation} \label{eqn:rstarstar}
r^{**} = (kT\rho^{-1}c_p{\varepsilon_T}^{\frac{5n}{4}-\frac{2}{3}}{\tau_p}^{\frac{15n}{4}})^\frac{1}{\frac{2}{3}+\frac{5n}{2}} .
\end{equation}
The polymer relaxation time $\tau_p \approx (N^{3/5}a)^3\mu/kT $,  where $N$ is the number of monomers per chain, $a$ is the length of a monomer, and $\mu$ is the dynamic viscosity of the fluid.

When writing down Eq.~\eqref{eqn:energy_balance}, de Gennes conjectured that the turbulence energy cascade is unaffected at scales $r > r^{**}$, which, however, is not consistent with the theory itself as it assumes that polymers already gain elastic energy from the stretching by eddies of size $r^{**} < r < r^{*}$ and hence must have diverted part of the turbulence energy flux at scales $r > r^{**}$. Assuming the time scale for polymers to transfer elastic energy down scales is $\tau_p$, then the elastic energy flux is 
\begin{equation}
\varepsilon_e(r) \sim E_e(r)/(\tau_p \rho) \sim \frac{kTc_p}{\tau_p\rho}(\frac{r^*}{r})^{5n/2} .
\label{eq:elas_energy_flux} 
\end{equation}
The elastic energy flux given above increases as $r$ decreases and  can dominate the turbulence energy flux $\varepsilon_T$, as suggested in Fig.~\ref{fig:balances}(b). A new scale $r_\varepsilon$ can be defined when the two fluxes balance:
\begin{equation} \label{eqn:energy_rate_balance}
\varepsilon_T = \varepsilon_e(r_\varepsilon) = A\frac{kTc_p}{\tau_p\rho}(\frac{r^*}{r_\varepsilon})^{5n/2},
\end{equation} 
which yields
\begin{equation} \label{eqn:r_epsilon}
r_{\varepsilon} = (AkT/\rho)^{\frac{2}{5n}}{c_p}^{\frac{2}{5n}}{\varepsilon_T}^{\frac{1}{2}-\frac{2}{5n}}{\tau_{p}}^{\frac{3}{2}-\frac{2}{5n}},
\end{equation}
where a proportional factor $A$ is introduced to convert the scaling relation to an equation.
For $r_\varepsilon < r < r^*$, the turbulent energy flux is only slightly affected and inertial range scaling of the cascade would still capture the behavior to leading order. However, for $r < r_\varepsilon$, the turbulent energy transfer will be strongly reduced. 

\begin{figure} [h]
\includegraphics[width=0.8\columnwidth]{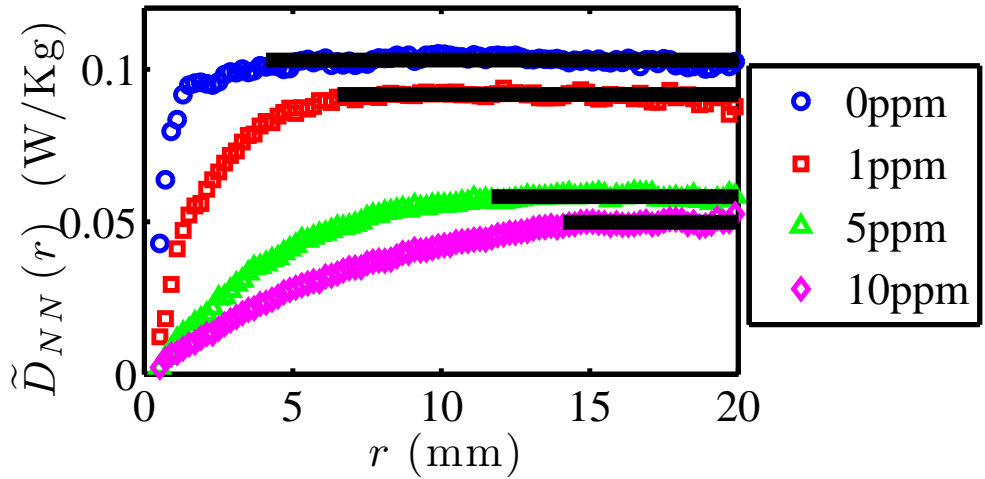}
\caption{\label{fig:compensatedDNN}  Compensated second order transverse velocity structure functions $\widetilde{D}_{NN}$ at $R_\lambda = 360$ with different polymer concentrations. }
\end{figure}

As a test of this conjecture we investigated experimentally the validity of the elastic energy flux as given by Eq.~\ref{eqn:energy_rate_balance}. The experiment was done in the von K\'arm\'an swirling (VKS) flow system (see Methods for details). We probed the energy transfer in the flow by measuring the second order transverse velocity structure function $D_{NN}(r)=\langle|u_\perp(\mathbf{x}+\mathbf{r})-u_\perp(\mathbf{x})|^2\rangle$, where $u_\perp$ is the velocity component in the direction perpendicular to the separation vector $\mathbf{r}$. For Newtonian fluids, in the inertial range,  $D_{NN}(r)=\frac{4}{3}C_2(\varepsilon_T r)^{2/3}$, where $C_2$ is a constant (here we use $C_2=2.11$). This expression can be rewritten as $\widetilde{D}_{NN} (r) = \frac{1}{r}[\frac{3}{4}D_{NN}(r)/C_2]^{3/2} = \varepsilon_T$. Figure~\ref{fig:compensatedDNN} shows $\widetilde{D}_{NN} (r)$ for the flows at fixed $R_\lambda = 360$ with different polymer concentrations. ($R_\lambda$ is a Reynolds number definition widely used in the turbulence community, which is related to the more familiar Reynolds number as $R_\lambda \sim Re^{1/2}$.) For the pure water case, $\widetilde{D}_{NN} (r)$ displays a plateau of value $\varepsilon_T$ in the inertial range. When polymers were added, $\widetilde{D}_{NN} (r)$ reaches a plateau at larger $r$ and the value of the plateau is lower. Both the suppression of small scales~\cite{bonn:1993,tong:1992,liberzon:2005,liberzon:2006,berti:2006,crawford:2008,ouellette:2009} and the decreasing of $\varepsilon_T$~\cite{ouellette:2009} have been observed before. In~\cite{ouellette:2009}, it was also noticed that $\varepsilon_T$ measured from inertial range exceeds $\varepsilon_D$ measured from the dissipation range by a large amount, which is consistent with the conjecture that at small scales it is the polymers that transfer part of the fluid's fluctuation energy by elasticity and hence reduce the turbulence energy flux. 
Note that in~\cite{ouellette:2009} a critical polymer concentration of $\phi_c = 7$ ppm was observed below which $\varepsilon_T$ was not affected by the presence of polymers in the flow with similar $R_\lambda$ in VKS1. In  VKS2, $\varepsilon_T$ was found to decrease slightly even at $\phi = 1$ ppm. This difference is most likely due to the change in the large-scale flow, as a consequence of the modifications in propeller size and the vane structure. 
As we will show next, when data measured from the two apparatuses are processed the same way, the results overlap with each other, which strongly supports the conjecture that the mechanism of elastic energy transfer by polymers is independent of how energy in injected into the flow at large scales.

\begin{figure}
\includegraphics[width=1\columnwidth]{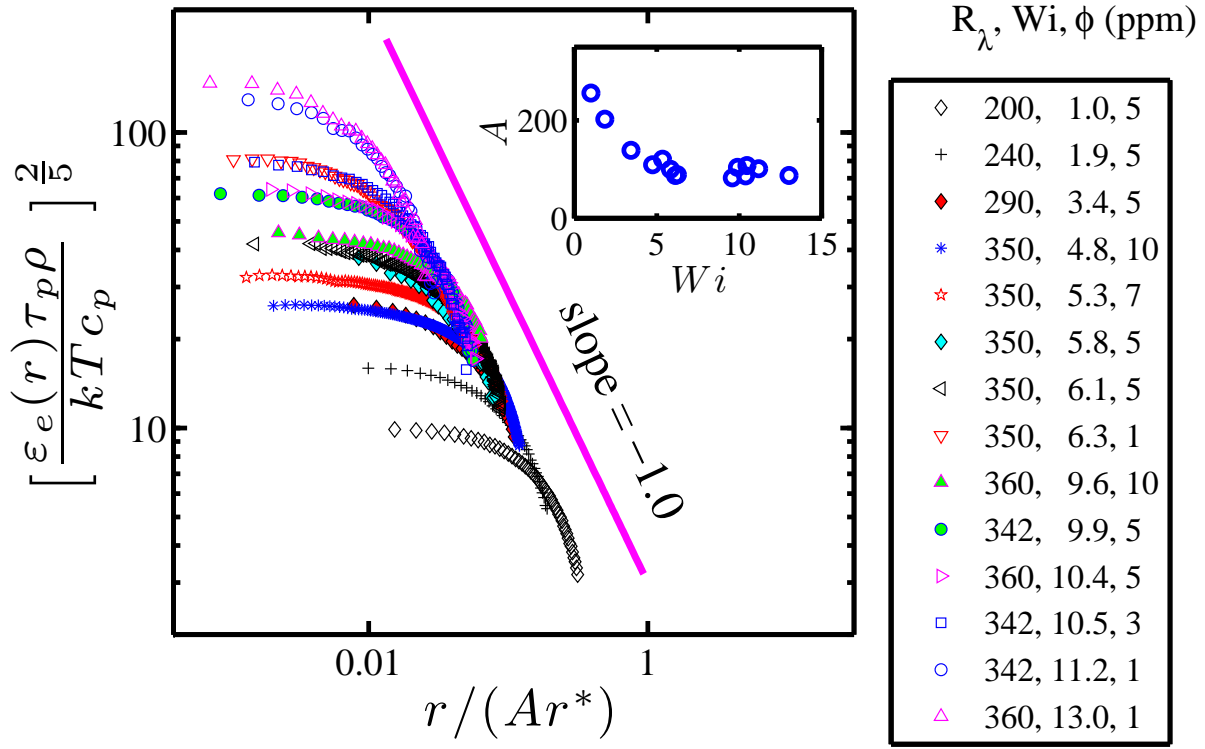}
\centering
\caption{\label{fig:elas_Energy_transfer} The measured elastic energy transfer by polymers $\varepsilon_e(r)$ in the form of $[\frac{\varepsilon_e(r)\tau_p\rho}{kTc_p}]^\frac{2}{5}$ vs. $r/(Ar^*)$ for 14 different data sets. Inset: the numerical proportional factor $A$ obtained from fitting each of the 14 data sets to $[\frac{\varepsilon_e(r)\tau_p\rho}{kTc_p}]^\frac{2}{5} = A(r/r^*)^{-1.0}$. }
\end{figure}

In principle, the difference in the two curves of $\widetilde{D}_{NN} (r)$ shown in Fig.~\ref{fig:compensatedDNN} gives the energy flux by polymers $\varepsilon_e(r)$. However, as shown in~\cite{ouellette:2009}, for VKS flows polymers reduce large scale velocity fluctuations also, which contributes to the decrease in $\varepsilon_T$. To account for this change we assume that the plateau values of $\widetilde{D}_{NN} (r)$ for polymer solutions correspond to the total energy transfer rate $\varepsilon_T$. The difference between $\varepsilon_T$ and $\widetilde{D}_{NN} (r)$ at smaller $r$ is thus $\varepsilon_e(r)$.  As we shall see later, this assumption is well supported by the observed collapse of the data. 
Moreover, we notice that Eq.~\eqref{eq:elas_energy_flux} can be rearranged as
\begin{equation} \label{eqn:elastic_epsilon_a}
\left[ \frac{{\varepsilon_e} (r) \tau_p\rho}{kTc_p} \right]^{2/5} = A(r^*/r)^n .
\end{equation}
 From our data we fit the constant $A$ and the exponent $n$ using the measured $[\frac{{\varepsilon_e} (r) \tau_p\rho}{kTc_p}]^{2/5}$ in the form of Eq.~\eqref{eqn:elastic_epsilon_a}. In the inertial range our data support the power-law relation for scales $r \gg \eta$, \textit{i.e.}, $[\frac{{\varepsilon_e} (r) \tau_p\rho}{kTc_p}]^{2/5} \sim (r/r^*)^n$. From the 14 data sets, we obtain $n=1.0\pm0.2$. As shown in Fig.~\ref{fig:elas_Energy_transfer}, all the 14 data sets are consistent with $(r/r^*)^{-1}$ scaling for relatively large $r$. In the inset of Fig.~\ref{fig:elas_Energy_transfer}, we show the values of $A$ fitted with $n=1$ from each data set as a function of $Wi$. Except for the two small $Wi$ cases, $A$ is within $101\pm17$ throughout the parameter ranges we explored. The larger values of $A$ found for the two low $Wi$ cases ($Wi=1.0$ and $1.9$) is very likely due to that the $Wi$ are below a critical $Wi_c$ recently found in  numerical simulation where $Wi_c$ is reported to be $3 \sim 4$~\cite{watanabe:2010} .

Previously some of us~\cite{ouellette:2009} had found that data from varying Reynolds number but constant polymer concentration ($R_\lambda = 200, 240, 290$, and $350$, $\phi = 5$ ppm) could be collapsed with the normalization scale $\sim Wi^{-0.58}$~\cite{ouellette:2009}, which is equivalent to $\sim {\varepsilon_T}^{-0.29}$ or $n=0.51$. The range of $Wi$ in this previous work was relatively small ($Wi=1.0, 1.9, 3.4$, and $5.8$)and the change in the value of $A$  (see inset of Fig.~\ref{fig:elas_Energy_transfer}) masked the analysis.
 
According to the conjecture by Tabor \& de Gennes, the exponent $n=1.0$ implies that polymers are most effectively stretched in locally biaxial extensional regions of the flow. This is consistent with the properties of velocity gradients in turbulent flows and previous findings of polymer behavior in turbulence. In homogeneous and isotropic three-dimensional turbulence, locally biaxial stretching is more probable than uniaxial stretching~\cite{ashurst:1987}. Moreover, recent simulations showed that polymer extensions are larger in the biaxial stretching regions of a channel flow~\cite{terrapon:2004} or a homogeneous shear flow~\cite{peters:2007}. Furthermore, we note that a previous compilation of drag-reduction data from different turbulent pipe flows reported a value $n\approx 2/3$~\cite{sreenivasan:2000} that is quite close to what we found here from velocity structure functions in bulk turbulence.

\begin{figure}
\includegraphics[width=0.9\columnwidth]{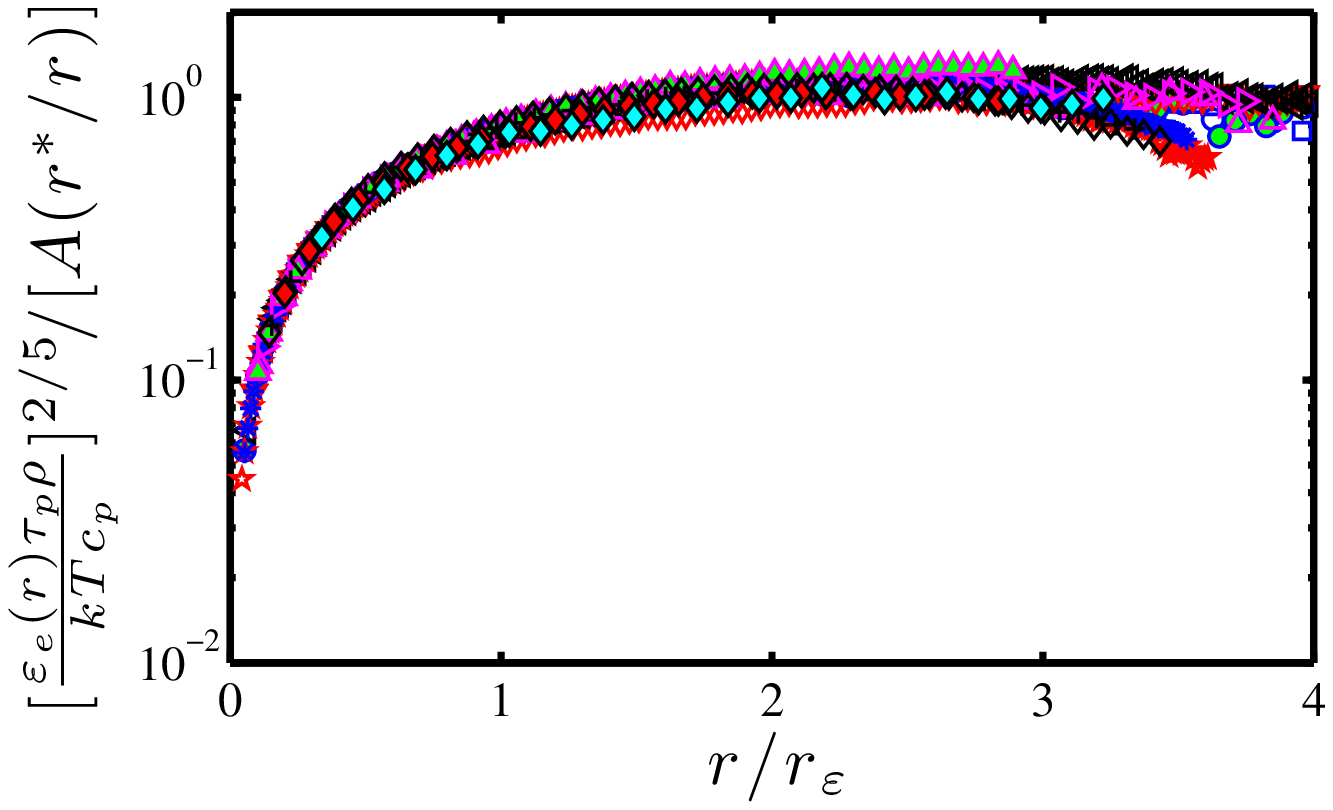}
\centering
\caption{\label{fig:epsilon_e_collapsed} Collapsing of the measured elastic energy transfer by polymers when $r$ is normalized by $r_\varepsilon$. In this compensated plot, ($[\frac{\varepsilon_e(r)\tau_p\rho}{kTc_p}]^\frac{2}{5}$ is divided by $A(r^*/r)^{1.0}$) and $r$ is rescaled by the elastic scale $r_\varepsilon$. The legend is the same as in Fig.~\ref{fig:elas_Energy_transfer}. }
\end{figure}

With $n=1$, we can determine the scale $r_{\varepsilon}$ from Eq.~\eqref{eqn:r_epsilon}:
\begin{equation}
r_{\varepsilon} =  A(kT/\rho)^{0.4}{c_p}^{0.4}{\varepsilon_T}^{0.1}{\tau_p}^{1.1} .
\end{equation}
In Fig.~\ref{fig:epsilon_e_collapsed} we plot $[\frac{\varepsilon_e(r)\tau_p\rho}{kTc_p}]^\frac{2}{5}/[A(r^*/r)^{1.0}]$ against $r/r_\varepsilon$. All the 14 data sets collapse and show a plateau when $r \gtrsim r_{\varepsilon}$, which is the expected scaling range for elastic energy flux. For $r<r_\varepsilon$ polymer elasticity dominates the energy transfer process, and the traditional K41 energy cascade does not hold anymore. Thus at those scales the elastic energy transfer rate could not be obtained by simply subtracting $\widetilde{D}_{NN} (r)$ for $\phi > 0$ from the $\phi=0$ case.

\begin{figure}
\centering
\subfigure[]{
\label{fig:DNN_a} 
\includegraphics[width=0.8\columnwidth]{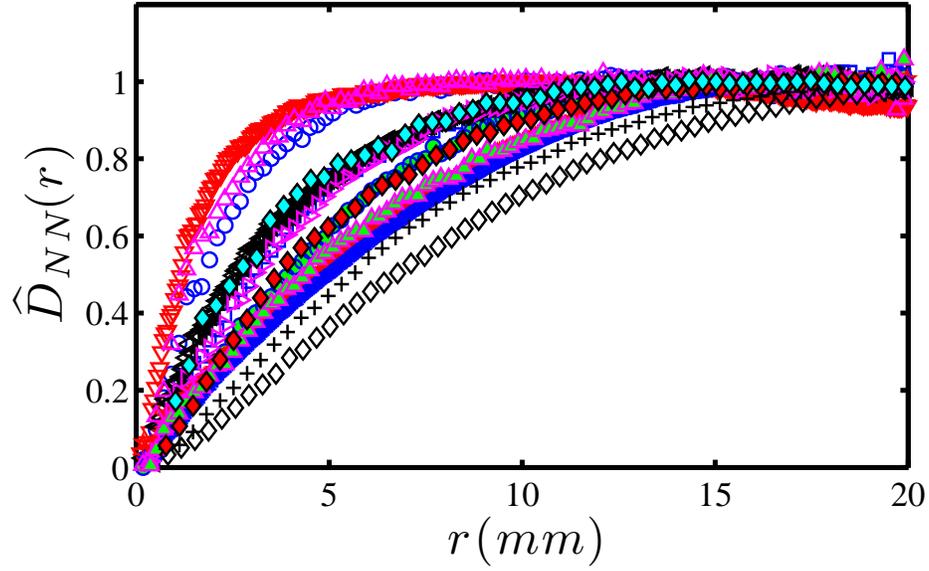}}
\subfigure[]{
\label{fig:DNN_b} 
\includegraphics[width=0.8\columnwidth]{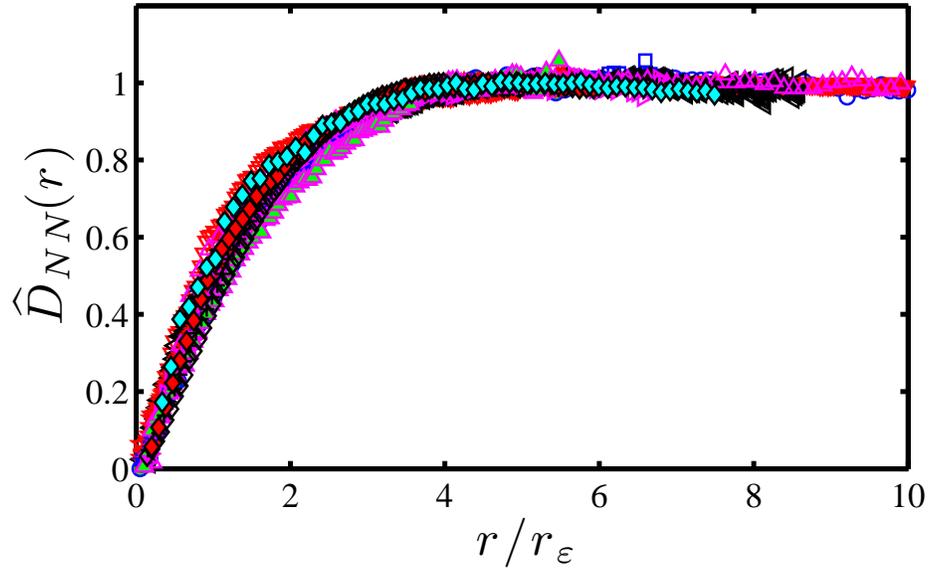}}
\centering
\caption{\label{fig:DNN} (a) The normalized second order velocity structure function $\widehat{D}_{NN}$ as functions of $r$ (a) and $r/r_\varepsilon$ (b) for the 14 data sets. The legend is the same as in Fig.~\ref{fig:elas_Energy_transfer}. }
\end{figure}

In addition to collapsing the elastic energy flux, $r_\varepsilon$ should also collapse the structure function data since $r_\varepsilon$ is the scale at which polymer elasticity dominates the turbulent energy cascade in the inertial range. Figure~\ref{fig:DNN} (a) shows $\widehat{D}_{NN}(r)$, which is defined as $\widetilde{D}_{NN}$ normalized by the corresponding $\varepsilon_T$. Clearly the lower ends of the inertial range are pushed to larger scales when polymers are present, because at the smaller scales the polymers are stretched and truncate the K41 type energy cascade. When $r$ is normalized by $r_{\varepsilon}$, $\widehat{D}_{NN}(r)$ from different data sets collapse, as shown in Fig. 5(b). We also performed experiments with a different polymer -- Polyethylene Oxide (PEO, $MW = 8\times 10^6$). The measured $\widehat{D}_{NN}$ at $R_\lambda = 342$  for two different concentrations ($5ppm$ and $10ppm$) are found to collapse with the PAM data but with a different value $A \approx 25$, which is consistent with the theory as the numerical factor $A$ depends on the combination of polymer and solvent. 

In a certain sense, $r_\varepsilon$ is similar to the Kolmogorov scale $\eta$ as both correspond to a scale at which the inertial range turbulent energy cascade is truncated by a mechanism whose effect is negligible in the inertial range but increases at small scales. On the other hand, $\eta$ for Newtonian flows and $r_\varepsilon$ for polymer solutions are significantly different. For Newtonian flows, the velocity field below $\eta$ is smooth and the small scale turbulence is expected to be universal. For polymer solutions, at scales $r < r_\varepsilon$ there is still room for interesting dynamics. It has been shown that in a smooth velocity field the elastic instability can drive polymer solutions to elastic turbulence~\cite{groisman:2000}. Our theory suggests that in turbulence this might occur when the polymer stress dominates the fluid stress, i.e., at scales below the de Gennes scale $r^{**}$. Numerical simulations of decaying turbulence with polymers indeed showed an enhancement of energy spectra at small scales (smaller than $\eta$)~\cite{perlekar:2006}. It will be very interesting to study the small scales experimentally with suitable diagnostic techniques.

\vspace{1cm}
\textbf{{\Large Methods}} \\

\textit{Elastic energy of polymer in turbulent flow} The behavior of a long-chain polymer is similar to that of a spring, except that the restoring force for the polymer chain is the entropic force. The elastic energy in one chain is $\sim kT[\lambda (r)]^{5/2} $, where $kT$ is the ``spring constant'' of the polymer and $\lambda (r)$ is the extension of the polymer due to eddies of size $r$. The polymer elastic energy is proportional to $\lambda^{5/2}$ instead of  $\lambda^{2}$ for a harmonic spring because of additional repulsions between monomers~\cite{pincus:1976}. The elastic energy per unit volume is then $\sim c_pkT\lambda^{5/2}$.
To formulate the polymer extension in a turbulent flow, de Gennes \& Tabor~\cite{tabor:1986,degennes:1986} considered the polymer extension in a laminar converging flow, where the flow velocity and shear rate are $1/(4\pi r^{D-1})$ and $1/(2\pi r^{D})$ respectively, where $D$ is the spatial dimension of the converging flow: $D=3$ means flow into a capillary tube and $D=2$ means flow into a long slit. The polymer starts to be stretched when the shear rate exceeds its relaxation $1/\tau_p$. Thus $1/(2\pi r^D) = 1/\tau_p$ gives a scale $r^*$: the polymer will be stretched when it is in the range $r<r^*$. The extension of the polymer is $\lambda = (r^*/r)^{D-1}$ since the polymer extends passively (affinely) with the deformation of the local volume element~\cite{daoudi:1978}. de Gennes \& Tabor then transferred this idea to the case of turbulent flows and assumed that the polymer extension can be written as $\lambda = (r^*/r)^n$, where $r$ is now the eddy size and $n$ is an unknown parameter given by the average stretching dimensions of the local flow field. 

\textit{Experimental setup and techniques} The turbulent flows were generated by two counter-rotating baffled disks in a cylindrical tank, known as the von K\'arm\'an swirling flow. Two VKS apparatuses were used in the experiments. One has been described in details before~\cite{voth:2002}. The diameter of the disk was $25$ cm. The diameter and the height of the tank were $49$ cm  and $63$ cm, respectively. The other VKS had nearly exactly the same  but with slightly smaller disks ($20$ cm of radius). The vane inserts used to break the large scale flow in the tank were also slightly different. Using Lagrangian particle tracking technique\cite{ouellette:2006a,xu:2008b}, we tracked simultaneously hundreds of tracer particles in a measurement volume of approximately ($3$ cm)$^3$ at the center of the flow, where the turbulence is close to homogenous and isotropic and the fluctuating velocities are much larger than the mean. The velocities of the particles were then obtained by differentiating the particle trajectories. The polymer used was  polyacrylamide (PAM, molecular weight $18\times 106$, Polysciences Inc.). In the experiments, $R_\lambda \equiv (15 u^4 /\varepsilon \nu)^{1/2}$ was varied between 200 and 360, $Wi \equiv \tau_p/\tau_\eta$ from 1.0 to 13.0, and polymer concentration $\phi$ from 0 to 10 ppm (parts per million by weight).  For the experiments with polymer solutions, we used the same $R_\lambda$ as that measured from the pure water case ($\phi = 0$) at the same rotating frequency of the propellers.

\vspace{1cm}
\textbf{{\Large Acknowledgment}} \\
We are grateful to the Max Planck Society and the Deutsche Forschungsgemeinschaft (through grant XU91-3) for their support. H.-D. Xi also thanks the Alexander von Humboldt Foundation for the generous support. We thank N. T. Ouellette for his contribution in the early stage of this study,  G. Ahlers and D. Funfschilling for useful discussions.

\newpage

\begin{thebibliography}{27}%
\makeatletter
\providecommand \@ifxundefined [1]{%
 \@ifx{#1\undefined}
}%
\providecommand \@ifnum [1]{%
 \ifnum #1\expandafter \@firstoftwo
 \else \expandafter \@secondoftwo
 \fi
}%
\providecommand \@ifx [1]{%
 \ifx #1\expandafter \@firstoftwo
 \else \expandafter \@secondoftwo
 \fi
}%
\providecommand \natexlab [1]{#1}%
\providecommand \enquote  [1]{``#1''}%
\providecommand \bibnamefont  [1]{#1}%
\providecommand \bibfnamefont [1]{#1}%
\providecommand \citenamefont [1]{#1}%
\providecommand \href@noop [0]{\@secondoftwo}%
\providecommand \href [0]{\begingroup \@sanitize@url \@href}%
\providecommand \@href[1]{\@@startlink{#1}\@@href}%
\providecommand \@@href[1]{\endgroup#1\@@endlink}%
\providecommand \@sanitize@url [0]{\catcode `\\12\catcode `\$12\catcode
  `\&12\catcode `\#12\catcode `\^12\catcode `\_12\catcode `\%12\relax}%
\providecommand \@@startlink[1]{}%
\providecommand \@@endlink[0]{}%
\providecommand \url  [0]{\begingroup\@sanitize@url \@url }%
\providecommand \@url [1]{\endgroup\@href {#1}{\urlprefix }}%
\providecommand \urlprefix  [0]{URL }%
\providecommand \Eprint [0]{\href }%
\providecommand \doibase [0]{http://dx.doi.org/}%
\providecommand \selectlanguage [0]{\@gobble}%
\providecommand \bibinfo  [0]{\@secondoftwo}%
\providecommand \bibfield  [0]{\@secondoftwo}%
\providecommand \translation [1]{[#1]}%
\providecommand \BibitemOpen [0]{}%
\providecommand \bibitemStop [0]{}%
\providecommand \bibitemNoStop [0]{.\EOS\space}%
\providecommand \EOS [0]{\spacefactor3000\relax}%
\providecommand \BibitemShut  [1]{\csname bibitem#1\endcsname}%
\let\auto@bib@innerbib\@empty
\bibitem [{\citenamefont {Richardson}(1922)}]{richardson:1922}%
  \BibitemOpen
  \bibfield  {author} {\bibinfo {author} {\bibfnamefont {L.~F.}\ \bibnamefont
  {Richardson}},\ }\href@noop {} {\emph {\bibinfo {title} {Weather Prediction
  by Numerical Process}}}\ (\bibinfo  {publisher} {Cambridge University
  Press},\ \bibinfo {address} {Cambridge, England},\ \bibinfo {year}
  {1922})\BibitemShut {NoStop}%
\bibitem [{\citenamefont {Kolmogorov}(1941)}]{K41}%
  \BibitemOpen
  \bibfield  {author} {\bibinfo {author} {\bibfnamefont {A.~N.}\ \bibnamefont
  {Kolmogorov}},\ }\bibfield  {title} {\enquote {\bibinfo {title} {The local
  structure of turbulence in incompressible viscous fluid for very large
  {R}eynolds numbers},}\ }\href@noop {} {\bibfield  {journal} {\bibinfo
  {journal} {Dokl. Akad. Nauk SSSR}\ }\textbf {\bibinfo {volume} {30}},\
  \bibinfo {pages} {301--305} (\bibinfo {year} {1941})}\BibitemShut {NoStop}%
\bibitem [{\citenamefont {Frisch}(1995)}]{frisch:1995}%
  \BibitemOpen
  \bibfield  {author} {\bibinfo {author} {\bibfnamefont {U.}~\bibnamefont
  {Frisch}},\ }\href@noop {} {\emph {\bibinfo {title} {Turbulence: The Legacy
  of A.~N.~Kolmogorov}}}\ (\bibinfo  {publisher} {Cambridge University Press},\
  \bibinfo {address} {Cambridge, England},\ \bibinfo {year} {1995})\BibitemShut
  {NoStop}%
\bibitem [{\citenamefont {Groisman}\ and\ \citenamefont
  {Steinberg}(2000)}]{groisman:2000}%
  \BibitemOpen
  \bibfield  {author} {\bibinfo {author} {\bibfnamefont {A.}~\bibnamefont
  {Groisman}}\ and\ \bibinfo {author} {\bibfnamefont {V.}~\bibnamefont
  {Steinberg}},\ }\bibfield  {title} {\enquote {\bibinfo {title} {Elastic
  turbulence in polymer solution flow},}\ }\href@noop {} {\bibfield  {journal}
  {\bibinfo  {journal} {Nature}\ }\textbf {\bibinfo {volume} {405}},\ \bibinfo
  {pages} {53--55} (\bibinfo {year} {2000})}\BibitemShut {NoStop}%
\bibitem [{\citenamefont {Procaccia}\ \emph {et~al.}(2008)\citenamefont
  {Procaccia}, \citenamefont {L'vov},\ and\ \citenamefont
  {Benzi}}]{procaccia:2008}%
  \BibitemOpen
  \bibfield  {author} {\bibinfo {author} {\bibfnamefont {I.}~\bibnamefont
  {Procaccia}}, \bibinfo {author} {\bibfnamefont {V.~S.}\ \bibnamefont
  {L'vov}}, \ and\ \bibinfo {author} {\bibfnamefont {R.}~\bibnamefont
  {Benzi}},\ }\bibfield  {title} {\enquote {\bibinfo {title} {Theory of drag
  reduction by polymer in wall-bounded turbulence},}\ }\href@noop {} {\bibfield
   {journal} {\bibinfo  {journal} {Rev.~Mod.~Phys.}\ }\textbf {\bibinfo
  {volume} {80}},\ \bibinfo {pages} {225--247} (\bibinfo {year}
  {2008})}\BibitemShut {NoStop}%
\bibitem [{\citenamefont {Tabor}\ and\ \citenamefont {{de
  Gennes}}(1986)}]{tabor:1986}%
  \BibitemOpen
  \bibfield  {author} {\bibinfo {author} {\bibfnamefont {M.}~\bibnamefont
  {Tabor}}\ and\ \bibinfo {author} {\bibfnamefont {P.~G.}\ \bibnamefont {{de
  Gennes}}},\ }\bibfield  {title} {\enquote {\bibinfo {title} {A cascade theory
  of drag reduction},}\ }\href@noop {} {\bibfield  {journal} {\bibinfo
  {journal} {Europhys.~Lett.}\ }\textbf {\bibinfo {volume} {2}},\ \bibinfo
  {pages} {519--522} (\bibinfo {year} {1986})}\BibitemShut {NoStop}%
\bibitem [{\citenamefont {{de Gennes}}(1986)}]{degennes:1986}%
  \BibitemOpen
  \bibfield  {author} {\bibinfo {author} {\bibfnamefont {P.~G.}\ \bibnamefont
  {{de Gennes}}},\ }\bibfield  {title} {\enquote {\bibinfo {title} {Towards a
  scaling theory of drag reduction},}\ }\href@noop {} {\bibfield  {journal}
  {\bibinfo  {journal} {Physica A}\ }\textbf {\bibinfo {volume} {140}},\
  \bibinfo {pages} {9--25} (\bibinfo {year} {1986})}\BibitemShut {NoStop}%
\bibitem [{\citenamefont {Sreenivasan}\ and\ \citenamefont
  {White}(2000)}]{sreenivasan:2000}%
  \BibitemOpen
  \bibfield  {author} {\bibinfo {author} {\bibfnamefont {K.~R.}\ \bibnamefont
  {Sreenivasan}}\ and\ \bibinfo {author} {\bibfnamefont {C.~M.}\ \bibnamefont
  {White}},\ }\bibfield  {title} {\enquote {\bibinfo {title} {The onset of drag
  reduction by dilute polymer additives, and the maximum drag reduction
  asymptote},}\ }\href@noop {} {\bibfield  {journal} {\bibinfo  {journal}
  {J.~Fluid Mech.}\ }\textbf {\bibinfo {volume} {409}},\ \bibinfo {pages}
  {149--164} (\bibinfo {year} {2000})}\BibitemShut {NoStop}%
\bibitem [{\citenamefont {Ouellette}\ \emph {et~al.}(2009)\citenamefont
  {Ouellette}, \citenamefont {Xu},\ and\ \citenamefont
  {Bodenschatz}}]{ouellette:2009}%
  \BibitemOpen
  \bibfield  {author} {\bibinfo {author} {\bibfnamefont {N.~T.}\ \bibnamefont
  {Ouellette}}, \bibinfo {author} {\bibfnamefont {H.}~\bibnamefont {Xu}}, \
  and\ \bibinfo {author} {\bibfnamefont {E.}~\bibnamefont {Bodenschatz}},\
  }\bibfield  {title} {\enquote {\bibinfo {title} {Bulk turbulence in dilute
  polymer solutions},}\ }\href@noop {} {\bibfield  {journal} {\bibinfo
  {journal} {J.~Fluid Mech.}\ }\textbf {\bibinfo {volume} {629}},\ \bibinfo
  {pages} {375--385} (\bibinfo {year} {2009})}\BibitemShut {NoStop}%
\bibitem [{\citenamefont {Smith}\ \emph {et~al.}(1999)\citenamefont {Smith},
  \citenamefont {Babcock},\ and\ \citenamefont {Chu}}]{smith:1999}%
  \BibitemOpen
  \bibfield  {author} {\bibinfo {author} {\bibfnamefont {D.~E.}\ \bibnamefont
  {Smith}}, \bibinfo {author} {\bibfnamefont {H.~P.}\ \bibnamefont {Babcock}},
  \ and\ \bibinfo {author} {\bibfnamefont {S.}~\bibnamefont {Chu}},\ }\bibfield
   {title} {\enquote {\bibinfo {title} {Single-polymer dynamics in steady shear
  flow},}\ }\href@noop {} {\bibfield  {journal} {\bibinfo  {journal} {Science}\
  }\textbf {\bibinfo {volume} {283}},\ \bibinfo {pages} {1724--1727} (\bibinfo
  {year} {1999})}\BibitemShut {NoStop}%
\bibitem [{\citenamefont {Lumley}(1973)}]{lumley:1973}%
  \BibitemOpen
  \bibfield  {author} {\bibinfo {author} {\bibfnamefont {J.~L.}\ \bibnamefont
  {Lumley}},\ }\bibfield  {title} {\enquote {\bibinfo {title} {Drag reduction
  in turbulent flow by polymer additives},}\ }\href@noop {} {\bibfield
  {journal} {\bibinfo  {journal} {J.~Polymer Sci.}\ }\textbf {\bibinfo {volume}
  {7}},\ \bibinfo {pages} {263--290} (\bibinfo {year} {1973})}\BibitemShut
  {NoStop}%
\bibitem [{\citenamefont {Bonn}\ \emph {et~al.}(1993)\citenamefont {Bonn},
  \citenamefont {Couder}, \citenamefont {van Dam},\ and\ \citenamefont
  {Douady}}]{bonn:1993}%
  \BibitemOpen
  \bibfield  {author} {\bibinfo {author} {\bibfnamefont {D.}~\bibnamefont
  {Bonn}}, \bibinfo {author} {\bibfnamefont {Y.}~\bibnamefont {Couder}},
  \bibinfo {author} {\bibfnamefont {P.~H.~J.}\ \bibnamefont {van Dam}}, \ and\
  \bibinfo {author} {\bibfnamefont {S.}~\bibnamefont {Douady}},\ }\bibfield
  {title} {\enquote {\bibinfo {title} {From small scales to large scales in
  three-dimensional turbulence: The effect of diluted polymers},}\ }\href@noop
  {} {\bibfield  {journal} {\bibinfo  {journal} {Phys. Rev. E}\ }\textbf
  {\bibinfo {volume} {47}},\ \bibinfo {pages} {R28} (\bibinfo {year}
  {1993})}\BibitemShut {NoStop}%
\bibitem [{\citenamefont {Tong}\ \emph {et~al.}(1992)\citenamefont {Tong},
  \citenamefont {Goldburg},\ and\ \citenamefont {Huang}}]{tong:1992}%
  \BibitemOpen
  \bibfield  {author} {\bibinfo {author} {\bibfnamefont {P.}~\bibnamefont
  {Tong}}, \bibinfo {author} {\bibfnamefont {W.~I.}\ \bibnamefont {Goldburg}},
  \ and\ \bibinfo {author} {\bibfnamefont {J.~S.}\ \bibnamefont {Huang}},\
  }\bibfield  {title} {\enquote {\bibinfo {title} {Measured effects of polymer
  additives on turbulent-velocity fluctuations at various length scales},}\
  }\href@noop {} {\bibfield  {journal} {\bibinfo  {journal} {Phy. Rev. A}\
  }\textbf {\bibinfo {volume} {45}},\ \bibinfo {pages} {7231} (\bibinfo {year}
  {1992})}\BibitemShut {NoStop}%
\bibitem [{\citenamefont {Liberzon}\ \emph {et~al.}(2005)\citenamefont
  {Liberzon}, \citenamefont {Guala}, \citenamefont {L\"uthi}, \citenamefont
  {Kinzelbach},\ and\ \citenamefont {Tsinober}}]{liberzon:2005}%
  \BibitemOpen
  \bibfield  {author} {\bibinfo {author} {\bibfnamefont {A.}~\bibnamefont
  {Liberzon}}, \bibinfo {author} {\bibfnamefont {M.}~\bibnamefont {Guala}},
  \bibinfo {author} {\bibfnamefont {B.}~\bibnamefont {L\"uthi}}, \bibinfo
  {author} {\bibfnamefont {W.}~\bibnamefont {Kinzelbach}}, \ and\ \bibinfo
  {author} {\bibfnamefont {A.}~\bibnamefont {Tsinober}},\ }\bibfield  {title}
  {\enquote {\bibinfo {title} {Turbulence in dilute polymer solutions},}\
  }\href@noop {} {\bibfield  {journal} {\bibinfo  {journal} {Phys.~Fluids}\
  }\textbf {\bibinfo {volume} {17}},\ \bibinfo {pages} {031707} (\bibinfo
  {year} {2005})}\BibitemShut {NoStop}%
\bibitem [{\citenamefont {Liberzon}\ \emph {et~al.}(2006)\citenamefont
  {Liberzon}, \citenamefont {Guala}, \citenamefont {Kinzelbach},\ and\
  \citenamefont {Tsinober}}]{liberzon:2006}%
  \BibitemOpen
  \bibfield  {author} {\bibinfo {author} {\bibfnamefont {A.}~\bibnamefont
  {Liberzon}}, \bibinfo {author} {\bibfnamefont {M.}~\bibnamefont {Guala}},
  \bibinfo {author} {\bibfnamefont {W.}~\bibnamefont {Kinzelbach}}, \ and\
  \bibinfo {author} {\bibfnamefont {A.}~\bibnamefont {Tsinober}},\ }\bibfield
  {title} {\enquote {\bibinfo {title} {On turbulent kinetic energy production
  and dissipation in dilute polymer solutions},}\ }\href@noop {} {\bibfield
  {journal} {\bibinfo  {journal} {Phys.~Fluids}\ }\textbf {\bibinfo {volume}
  {18}},\ \bibinfo {pages} {125101} (\bibinfo {year} {2006})}\BibitemShut
  {NoStop}%
\bibitem [{\citenamefont {Berti}\ \emph {et~al.}(2006)\citenamefont {Berti},
  \citenamefont {Bistagnino}, \citenamefont {Boffetta}, \citenamefont
  {Celani},\ and\ \citenamefont {Musacchio}}]{berti:2006}%
  \BibitemOpen
  \bibfield  {author} {\bibinfo {author} {\bibfnamefont {S.}~\bibnamefont
  {Berti}}, \bibinfo {author} {\bibfnamefont {A.}~\bibnamefont {Bistagnino}},
  \bibinfo {author} {\bibfnamefont {G.}~\bibnamefont {Boffetta}}, \bibinfo
  {author} {\bibfnamefont {A.}~\bibnamefont {Celani}}, \ and\ \bibinfo {author}
  {\bibfnamefont {S.}~\bibnamefont {Musacchio}},\ }\bibfield  {title} {\enquote
  {\bibinfo {title} {Small-scale statistics of viscoelastic turbulence},}\
  }\href@noop {} {\bibfield  {journal} {\bibinfo  {journal} {Europhysics
  Lett.}\ }\textbf {\bibinfo {volume} {76(1)}},\ \bibinfo {pages} {63--69}
  (\bibinfo {year} {2006})}\BibitemShut {NoStop}%
\bibitem [{\citenamefont {Crawford}\ \emph {et~al.}(2008)\citenamefont
  {Crawford}, \citenamefont {Mordant}, \citenamefont {Xu},\ and\ \citenamefont
  {Bodenschatz}}]{crawford:2008}%
  \BibitemOpen
  \bibfield  {author} {\bibinfo {author} {\bibfnamefont {A.~M.}\ \bibnamefont
  {Crawford}}, \bibinfo {author} {\bibfnamefont {N.}~\bibnamefont {Mordant}},
  \bibinfo {author} {\bibfnamefont {H.}~\bibnamefont {Xu}}, \ and\ \bibinfo
  {author} {\bibfnamefont {E.}~\bibnamefont {Bodenschatz}},\ }\bibfield
  {title} {\enquote {\bibinfo {title} {Fluid acceleration in the bulk of
  turbulent dilute polymer solutions},}\ }\href@noop {} {\bibfield  {journal}
  {\bibinfo  {journal} {New J.~Phys.}\ }\textbf {\bibinfo {volume} {10}},\
  \bibinfo {pages} {123015} (\bibinfo {year} {2008})}\BibitemShut {NoStop}%
\bibitem [{\citenamefont {Watanabe}\ and\ \citenamefont
  {Gotoh}(2010)}]{watanabe:2010}%
  \BibitemOpen
  \bibfield  {author} {\bibinfo {author} {\bibfnamefont {T.}~\bibnamefont
  {Watanabe}}\ and\ \bibinfo {author} {\bibfnamefont {T.}~\bibnamefont
  {Gotoh}},\ }\bibfield  {title} {\enquote {\bibinfo {title} {Coil-stretch
  transition in an ensemble of polymers in isotropic turbulence},}\ }\href@noop
  {} {\bibfield  {journal} {\bibinfo  {journal} {Phys.~Rev.~E}\ }\textbf
  {\bibinfo {volume} {81}},\ \bibinfo {pages} {066301} (\bibinfo {year}
  {2010})}\BibitemShut {NoStop}%
\bibitem [{\citenamefont {Ashurst}\ \emph {et~al.}(1987)\citenamefont
  {Ashurst}, \citenamefont {Kerstein}, \citenamefont {Kerr},\ and\
  \citenamefont {Gibson}}]{ashurst:1987}%
  \BibitemOpen
  \bibfield  {author} {\bibinfo {author} {\bibfnamefont {W.~T.}\ \bibnamefont
  {Ashurst}}, \bibinfo {author} {\bibfnamefont {A.~R.}\ \bibnamefont
  {Kerstein}}, \bibinfo {author} {\bibfnamefont {R.~M.}\ \bibnamefont {Kerr}},
  \ and\ \bibinfo {author} {\bibfnamefont {C.~H.}\ \bibnamefont {Gibson}},\
  }\bibfield  {title} {\enquote {\bibinfo {title} {Alignment of vorticity and
  scalar gradient with strain rate in simulated navier-stokes turbulence},}\
  }\href@noop {} {\bibfield  {journal} {\bibinfo  {journal} {Phys. Fluids}\
  }\textbf {\bibinfo {volume} {30}},\ \bibinfo {pages} {2343--2353} (\bibinfo
  {year} {1987})}\BibitemShut {NoStop}%
\bibitem [{\citenamefont {Terrapon}\ \emph {et~al.}(2004)\citenamefont
  {Terrapon}, \citenamefont {Dubief}, \citenamefont {Moin}, \citenamefont
  {Shaqfeh},\ and\ \citenamefont {Lele}}]{terrapon:2004}%
  \BibitemOpen
  \bibfield  {author} {\bibinfo {author} {\bibfnamefont {V.~E.}\ \bibnamefont
  {Terrapon}}, \bibinfo {author} {\bibfnamefont {Y.}~\bibnamefont {Dubief}},
  \bibinfo {author} {\bibfnamefont {P.}~\bibnamefont {Moin}}, \bibinfo {author}
  {\bibfnamefont {E.~S.~G.}\ \bibnamefont {Shaqfeh}}, \ and\ \bibinfo {author}
  {\bibfnamefont {S.~K.}\ \bibnamefont {Lele}},\ }\bibfield  {title} {\enquote
  {\bibinfo {title} {Simulated polymer stretch in a turbulent flow using
  brownian dynamics},}\ }\href@noop {} {\bibfield  {journal} {\bibinfo
  {journal} {J.~Fluid Mech.}\ }\textbf {\bibinfo {volume} {504}},\ \bibinfo
  {pages} {61} (\bibinfo {year} {2004})}\BibitemShut {NoStop}%
\bibitem [{\citenamefont {Peters}\ and\ \citenamefont
  {Schumacher}(2007)}]{peters:2007}%
  \BibitemOpen
  \bibfield  {author} {\bibinfo {author} {\bibfnamefont {T.}~\bibnamefont
  {Peters}}\ and\ \bibinfo {author} {\bibfnamefont {J.}~\bibnamefont
  {Schumacher}},\ }\bibfield  {title} {\enquote {\bibinfo {title} {Two-way
  coupling of finitely extensible nonlinear elastic dumbbells with a turbulent
  shear flow},}\ }\href@noop {} {\bibfield  {journal} {\bibinfo  {journal}
  {Phys.~Fluids}\ }\textbf {\bibinfo {volume} {19}},\ \bibinfo {pages} {065109}
  (\bibinfo {year} {2007})}\BibitemShut {NoStop}%
\bibitem [{\citenamefont {Perlekar}\ \emph {et~al.}(2006)\citenamefont
  {Perlekar}, \citenamefont {Mitra},\ and\ \citenamefont
  {Pandit}}]{perlekar:2006}%
  \BibitemOpen
  \bibfield  {author} {\bibinfo {author} {\bibfnamefont {P.}~\bibnamefont
  {Perlekar}}, \bibinfo {author} {\bibfnamefont {D.}~\bibnamefont {Mitra}}, \
  and\ \bibinfo {author} {\bibfnamefont {R.}~\bibnamefont {Pandit}},\
  }\bibfield  {title} {\enquote {\bibinfo {title} {Manifestations of drag
  reduction by polymer additives in decaying, homogeneous, isotropic
  turbulence},}\ }\href@noop {} {\bibfield  {journal} {\bibinfo  {journal}
  {Phys.~Rev.~Lett.}\ }\textbf {\bibinfo {volume} {97}},\ \bibinfo {pages}
  {264501} (\bibinfo {year} {2006})}\BibitemShut {NoStop}%
\bibitem [{\citenamefont {Pincus}(1976)}]{pincus:1976}%
  \BibitemOpen
  \bibfield  {author} {\bibinfo {author} {\bibfnamefont {P.}~\bibnamefont
  {Pincus}},\ }\bibfield  {title} {\enquote {\bibinfo {title} {Excluded volume
  effects and stretched polymer chains},}\ }\href@noop {} {\bibfield  {journal}
  {\bibinfo  {journal} {Macromolecules}\ }\textbf {\bibinfo {volume} {9}},\
  \bibinfo {pages} {386} (\bibinfo {year} {1976})}\BibitemShut {NoStop}%
\bibitem [{\citenamefont {Daoudi}\ and\ \citenamefont
  {Brochard}(1978)}]{daoudi:1978}%
  \BibitemOpen
  \bibfield  {author} {\bibinfo {author} {\bibfnamefont {S.}~\bibnamefont
  {Daoudi}}\ and\ \bibinfo {author} {\bibfnamefont {F.}~\bibnamefont
  {Brochard}},\ }\bibfield  {title} {\enquote {\bibinfo {title} {Flows of
  flexible polymer solutions in pores},}\ }\href@noop {} {\bibfield  {journal}
  {\bibinfo  {journal} {Macromolecules}\ }\textbf {\bibinfo {volume} {11}},\
  \bibinfo {pages} {751} (\bibinfo {year} {1978})}\BibitemShut {NoStop}%
\bibitem [{\citenamefont {Voth}\ \emph {et~al.}(2002)\citenamefont {Voth},
  \citenamefont {La~Porta}, \citenamefont {Crawford}, \citenamefont
  {Alexander},\ and\ \citenamefont {Bodenschatz}}]{voth:2002}%
  \BibitemOpen
  \bibfield  {author} {\bibinfo {author} {\bibfnamefont {G.~A.}\ \bibnamefont
  {Voth}}, \bibinfo {author} {\bibfnamefont {A.}~\bibnamefont {La~Porta}},
  \bibinfo {author} {\bibfnamefont {A.~M.}\ \bibnamefont {Crawford}}, \bibinfo
  {author} {\bibfnamefont {J.}~\bibnamefont {Alexander}}, \ and\ \bibinfo
  {author} {\bibfnamefont {E.}~\bibnamefont {Bodenschatz}},\ }\bibfield
  {title} {\enquote {\bibinfo {title} {Measurement of particle accelerations in
  fully developed turbulence},}\ }\href@noop {} {\bibfield  {journal} {\bibinfo
   {journal} {J.~Fluid Mech.}\ }\textbf {\bibinfo {volume} {469}},\ \bibinfo
  {pages} {121--160} (\bibinfo {year} {2002})}\BibitemShut {NoStop}%
\bibitem [{\citenamefont {Ouellette}\ \emph {et~al.}(2006)\citenamefont
  {Ouellette}, \citenamefont {Xu},\ and\ \citenamefont
  {Bodenschatz}}]{ouellette:2006a}%
  \BibitemOpen
  \bibfield  {author} {\bibinfo {author} {\bibfnamefont {N.~T.}\ \bibnamefont
  {Ouellette}}, \bibinfo {author} {\bibfnamefont {H.}~\bibnamefont {Xu}}, \
  and\ \bibinfo {author} {\bibfnamefont {E.}~\bibnamefont {Bodenschatz}},\
  }\bibfield  {title} {\enquote {\bibinfo {title} {A quantitative study of
  three-dimensional {L}agrangian particle tracking algorithms},}\ }\href@noop
  {} {\bibfield  {journal} {\bibinfo  {journal} {Exp.~Fluids}\ }\textbf
  {\bibinfo {volume} {40}},\ \bibinfo {pages} {301--313} (\bibinfo {year}
  {2006})}\BibitemShut {NoStop}%
\bibitem [{\citenamefont {Xu}(2008)}]{xu:2008b}%
  \BibitemOpen
  \bibfield  {author} {\bibinfo {author} {\bibfnamefont {H.}~\bibnamefont
  {Xu}},\ }\bibfield  {title} {\enquote {\bibinfo {title} {Tracking
  {Lagrangian} trajectories in physical-velocity space},}\ }\href@noop {}
  {\bibfield  {journal} {\bibinfo  {journal} {Meas.~Sci.~Technol.}\ }\textbf
  {\bibinfo {volume} {19}},\ \bibinfo {pages} {075105} (\bibinfo {year}
  {2008})}\BibitemShut {NoStop}%
\end{thebibliography}
\end{document}